# Questions Concerning Schwarzschild's Solution of Einstein's Equations.

J. Dunning-Davies,
Department of Physics,
University of Hull,
Hull HU6 7RX;
England,
j.dunning-davies@hull.ac.uk

**Abstract.**

Now that an English translation of Schwarzschild's original work exists, that work has become accessible to more people. Here his original solution to the Einstein field equations is examined and it is noted that it does not contain the mathematical singularity normally associated with the existence of a black hole. Einstein's own views on this subject are considered also and it is seen that, at the very least, grave questions exist over the possible existence of these somewhat esoteric stellar objects.

## Introduction.

In many of the standard textbooks on the General Theory of Relativity [1], time is devoted to discussing Schwarzschild's solution of the Einstein field equations. Normally, this solution is stated as being either

$$ds^2 = \left\{1 - \frac{2Gm}{rc^2}\right\}c^2 dt^2 - \left\{1 - \frac{2Gm}{rc^2}\right\}^{-1} dr^2 - r^2\left(d\theta^2 + \sin^2\theta d\phi^2\right) \quad (1)$$

or more usually

$$ds^2 = \left\{1 - \frac{2m}{r}\right\}dt^2 - \left\{1 - \frac{2m}{r}\right\}^{-1} dr^2 - r^2\left(d\theta^2 + \sin^2\theta d\phi^2\right) \quad (2)$$

where the universal constant of gravitation, $G$, and the speed of light, $c$, have both been put equal to unity. Here $r$, $\theta$, and $\phi$ appear to be taken to be normal polar co-ordinates.

In the above expressions, a mathematical singularity is seen to occur when $r=0$, as might be expected for polar co-ordinates. However, due to the form of the coefficient of $dr^2$, it follows that a second mathematical singularity occurs when, in (1), $rc^2 = 2Gm$ or, in (2), $r = 2m$. The first singularity is regularly dismissed as being merely a property of polar co-ordinates and, therefore, of no physical significance. The second singularity, however, tends to have a physical interpretation attributed to it - namely that it is said to indicate the existence of a black hole. Somewhat ironically, as will be seen later, this is referred to as a Schwarzschild black hole. If this interpretation were valid, it would imply that, for an object of mass $m$ and radius $r$ to be a black hole, it would need to satisfy the inequality

$$m/r \geq c^2/2G = 6.7 \times 10^{26} \text{ kg/m} \quad (3)$$

Incidentally, it has always seemed fascinating to realise that this expression for the ratio of mass to radius is the same as that derived, using purely Newtonian mechanics, by Michell in 1784 for a material body having an escape speed equal to, or greater than, the speed of light [2].

As stated above, many modern texts quote one of equations (1) or (2) as the Schwarzschild solution of the Einstein field equations, but is this so? Recently, an English translation of Schwarzschild's article of 1916 [3], has appeared and this has made the original work accessible to many more people. For this the scientific community owes the translators, S. Antoci and A. Loinger, a tremendous debt of gratitude. It also enables the above question to be raised by more people.

## The Schwarzschild Solution.

An excellent discussion of the Schwarzschild solution and its derivation is provided in chapter eighteen of the little book on the General Theory of Relativity by Dirac [4]. Here it is presented in the form (2) above and $r$, $\theta$, and $\phi$ are quite clearly stated to be the usual polar co-ordinates. It is pointed out that the case being considered is that of a static, spherically symmetric field produced by a spherically symmetric body at rest. After the completion of the derivation, it is noted that the said solution holds only outside the surface of the body producing the field, where there is no matter and, hence, it holds fairly accurately outside the surface of a star.

The following chapter is then devoted to the topic of black holes. It is noted that the Schwarzschild solution (2) becomes singular when $r = 2m$ and so it might appear that that value for $r$ indicated a minimum radius for a body of mass $m$ but it is claimed that a closer investigation reveals that this is not so. In the discussion which follows, the continuation of the Schwarzschild solution for values of $r < 2m$ is investigated. To achieve this, it is found necessary to use a non-static system of co-ordinates so that components of the metric tensor may vary with the time co-ordinate. This is accomplished by retaining $\theta$ and $\phi$ as co-ordinates but, instead of $t$ and $r$, using $\tau$ and $\rho$ defined by

$$\tau = t + f(r) \text{ and } \rho = t + g(r), \quad (4)$$

where the functions $f$ and $g$ are at the disposal of the investigator.

It transpires that, for the region $r < 2m$, the Schwarzschild solution is found to adopt the form

$$ds^2 = d\tau^2 - \frac{2m}{\mu(\rho-\tau)^{2/3}} d\rho^2 - \mu^2(\rho-\tau)^{4/3}\left(d\theta^2 + \sin^2\theta \, d\phi^2\right), \quad (5)$$

where $\mu = \left(\frac{3}{2}\sqrt{2m}\right)^{2/3}$. From the actual derivation, it follows that the critical value $r = 2m$ corresponds to $\rho - \tau = 4m/3$ and there is no singularity at this point in this metric.

From this point onwards, Dirac's argument becomes extremely interesting. He notes that the metric given by (5) satisfies Einstein's equations for empty space in the region $r > 2m$ because it may be transformed into the Schwarzschild solution by a simple change of co-ordinates. By analytic continuation, it is seen to satisfy the equations for $r \leq 2m$ also, because there is now no singularity at $r = 2m$. The singularity now appears, via equations (4), in the connection between old and new co-ordinates. Dirac then comments that, once the new co-ordinate system is established, the old one may be ignored and then the singularity appears no longer.

He comments further that the region of space for which $r > 2m$ may not communicate with that for which $r < 2m$. Also, any signal, even a light signal, would take an infinite time to cross the boundary at $r = 2m$. Thus, there can be no direct observational knowledge of the region for which $r < 2m$. If this argument were true, surely the region for which $r < 2m$ would lie outside our universe; would not really be a part of it? Dirac calls the region for which $r < 2m$ a black hole but is this an object in our physical three-dimensional space or one in an abstract, four-dimensional, mathematical space-time?

Finally, Dirac asks whether such a region exists and notes that the only definite statement which may be made is that the Einstein equations allow it. This is a question which will be considered further shortly but suffice it to say at this juncture that Einstein himself did not accept that it existed physically [5]. It is noted that a massive stellar object may collapse to an extremely small radius where the forces of gravity might become so strong that no known physical forces could withstand them and prevent further collapse. Such a situation would herald the collapse to a black hole but, as measured by our clocks, the final state would be achieved only after an infinite time. This argument would appear to stem from the ideas of Oppenheimer and Snyder [6]. They predicted that, when all sources of thermonuclear energy were exhausted, a large enough star would collapse and the contraction would continue indefinitely unless the star was able to reduce its mass sufficiently by some means. They also made the point that the total time for such a collapse would be finite for an observer co-moving with the stellar matter, although it would appear to take an infinite time for a distant observer. This was taken to indicate that the star tended to 'close itself off from any communication with a distant observer'; only its gravitational field persisting. Accepting this argument as valid for the moment, it might be asked, if such an object existed, would it *ever* be detectable by an external observer? On the other hand, if its gravitational field persists, and presumably the effects of that gravitational field on the surroundings, then, in a sense, the star is retaining some contact, albeit indirect, with a distant observer.

Also, for very many years, it has been noted that the transformation

$$\tau = t + u + \log(r - 2m)$$

applied to the Schwarzschild solution in the form (2) would remove the offending singularity. This was taken to indicate that the singularity was mathematical, ***not*** physical. This conclusion agrees with that of Einstein himself who, in an article of 1939 [5], concluded that the result of the investigation contained in that paper was a 'clear understanding as to why the "Schwarzschild singularities" do not exist in physical reality'. He went on to point out that, his investigation dealt only with clusters whose particles moved along circular paths but he felt it not unreasonable to feel that more general cases would have analogous results. He then stated quite categorically that 'the "Schwarzschild singularity" does not appear for the reason that matter cannot be concentrated arbitrarily'. This seems a very definite rejection of the notion of black holes by the very man often heralded as their father. If the general tone of his book is an indication of his view, then it seems to be the case that Dirac agreed with this interpretation also. This point concerning a possible physical interpretation of a mathematical singularity has been raised previously by Loinger [7], who has published a number of articles on arXiv in which the non-existence of black holes has been claimed. However, what of Schwarzschild himself; it's his solution of Einstein's equations which is really at the heart of this matter?

**Schwarzschild's Original Solution.**

As noted earlier, the translation of Schwarzschild's paper of 1916 [3] into English has made his work accessible to many more people. In his article, everything is written initially in terms of variables denoted by $x_1, x_2, x_3, x_4$ and the point is made that the field equations 'have the fundamental property that they preserve their form under the substitution of other arbitrary variables as long as the determinant of the substitution equals one'. The first three of the above co-ordinates are then taken to stand for rectangular co-ordinates, and the fourth is taken to be time. If these are denoted by $x, y, z,$ and $t$ the most general acceptable line element is then stated, but it is noted immediately that, when one goes over to polar co-ordinates according to the usual rules, the determinant of the transformation is not one. Hence, the field equations would not remain unaltered. Schwarzschild then employs the trick of putting

$$x_1 = r^3/3, x_2 = -\cos\theta, x_3 = \phi,$$

where $r$, $\theta$, $\phi$ are the normal polar co-ordinates. These new variables are then polar co-ordinates but with a determinant of the transformation equal to one. Schwarzschild then proceeds to derive his solution and presents it in the form

$$ds^2 = (1 - \alpha/R)dt^2 - (1 - \alpha/R)^{-1} dR^2 - R^2(d\theta^2 + \sin^2\theta d\phi^2),$$

where $R = (r^3 + \alpha^3)^{1/3}$.

Hence, Schwarzschild's actual solution does contain a singularity when $R = \alpha$, but $R$ is not the polar co-ordinate. It is clearly seen from above that, when $R = \alpha$, $r = 0$; that is, the singularity actually occurs at the origin of polar co-ordinates, as is usual. Therefore, according to Schwarzschild's own writing there is simply no singularity at $r = 2m$, to use the modern notation, and so the argument for general relativity predicting the existence of black holes cannot be justified by reference to the so-called Schwarzschild solution and it seems not a little ironic that non-rotating, uncharged black holes should be called Schwarzschild black holes.

**Conclusions.**

These days, claims for the identification of black holes appear fairly regularly in the scientific literature. Quite often, the supposed existence of black holes - even that of so-called massive black holes - is invoked to explain some otherwise puzzling phenomenon. However, so far, on no occasion has the postulated object satisfied the requirement mentioned earlier that, for a black hole, the ratio of the body's mass to its radius - or more specifically in general relativistic language, the radius of its event horizon - must be subject to the restriction

$$m/r \geq 6.7 \times 10^{26} \text{kg/m}$$

[8]. Now it emerges that the mathematical singularity at the centre of the discussion simply did not appear in Schwarzschild's original solution of Einstein's equations. Obviously mathematics was used by Schwarzschild to find this solution, but it was used meticulously. It was noted carefully that, if a transformation of coordinates for which the determinant of the transformation does not equal unity, is used, then the field equations themselves would not remain in an unaltered form. Hence, Schwarzschild adopted a transformation for which the value of the said determinant was one and went on to derive an exact, - not approximate, - solution to the equations. Also, Einstein himself proved that the singularity appearing in the popular form of the Schwarzschild solution has no physical significance. In all that Schwarzschild and Einstein did on this topic, the mathematics was a tool to help them achieve what they wanted. At no point was physical reality modified to fit a mathematical conclusion. This is the way things should be and provides an object lesson to many; - the mathematics is a tool and, as such, must be subservient to the physics.

Where then does that leave the modern notion of a black hole? Considerations such as those above, undoubtedly raise major questions about the basis of much modern work. The idea of a body being so dense that it's escape speed is greater than the speed of light remains not unreasonable though but, if the speed of light is a variable quantity - proportional, for example, to the square root of the background temperature, as suggested by Thornhill[9], Moffatt[10] and, more recently, Albrecht and Magueijo [11] - many new and interesting questions arise.

**References.**

[1] Adler, R., Bazin, M. & Schiffer, M., 1965, *Introduction to General Relativity*, (McGraw-Hill, New York)

[2] Michell, J., 1784, Philos. Trans. R. Soc., **74**, 35

[3] Schwarzschild, K., 1916, Sitzungsberichte der Königlich Preussischen Akademie der Wissenschaften zu Berlin, Phys.-Math. Klasse, 189 (translation by S. Antoci & A. Loinger, arXiv:physics/9905030)

[4] Dirac, P.A.M., 1996, *General Theory of Relativity*, (Princeton University Press, Princeton, New Jersey)

[5] Einstein, A., 1939, Annals of Mathematics, **40**, 922

[6] Oppenheimer, J.R. & Snyder, H., 1939, Phys. Rev. **56**, 455

[7] Loinger, A., arXiv:physics/0402088

[8] Dunning-Davies, J., 2004, Science, **305**, 1238

[9] Thornhill, C.K., 1985, Speculations in Sci. & Tech., **8**, 263

[10] Moffatt, J., 1993, Int. J. Mod. Phys. D., **2**, 351; 1993, Found. Phys., **23**, 411

[11] Albrecht, A. & Magueijo, J., 1999, Phys. Rev. D., **59**, 043516